\documentclass[sigconf,nonacm]{acmart}
\AtBeginDocument{%
  }

\setcitestyle{numbers,sort&compress}    
\usepackage[sidenotes=false]{dtrt}
\usepackage{comment}
\usepackage{todonotes}
\usepackage{xurl}
\usepackage{booktabs} 
\usepackage{multirow}
\usepackage{float}
\usepackage{makecell}
\usepackage{cleveref}
\usepackage{siunitx}
\usepackage{glossaries}
\usepackage{tabularx}
\usepackage{array}
\usepackage{graphicx}
\usepackage{subcaption}
\usepackage{xcolor}
\crefname{appendix}{appendix}{appendices}
\Crefname{appendix}{Appendix}{Appendices}
\theoremstyle{acmdefinition}\newtheorem{assumption}{Assumption}
\newacronym{pbs}{PBS}{Proposer-Builder Separation}
\newacronym{ofc}{OFC}{Order Flow Composition}
\newacronym{bft}{BFT}{Byzantine Fault Tolerant}
\newacronym{pos}{PoS}{Proof-of-Stake}
\newacronym{pow}{PoW}{Proof-of-Work}
\newacronym{mev}{MEV}{Maximal Extractable Value}
\newacronym[shortplural=DEXes]{dex}{DEX}{Decentralized Exchange}
\newacronym[shortplural=CEXes]{cex}{CEX}{Centralized Exchange} 
\newacronym{amm}{AMM}{Automated Market Maker}
\newacronym{defi}{DeFi}{Decentralized Finance}
\newacronym{tradfi}{TradFi}{Traditional Finance}
\newacronym{pga}{PGA}{Priority Gas Auction}
\newacronym{tob}{ToB}{Top-of-Block}
\newacronym{bob}{BoB}{Body-of-Block}
\newacronym{eob}{EoB}{End-of-Block}
\newacronym{ofa}{OFA}{Order Flow Auction}
\newacronym{0xce91228789b57deb45e66ca10ff648385fe7093b}{0xCe91228789B57DEb45e66Ca10Ff648385fE7093b}{MEV Blocker Rebates Safe}
\newacronym{rfq}{RFQ}{Request For Quote}
\newacronym{pm}{PM}{Profit Margin}
\newacronym{eoa}{EOA}{Externally Owned Account}
\newacronym{eof}{EOF}{Exclusive Order Flow}
\newacronym{msof}{MSOF}{Most Significant Order Flow}
\newacronym{jit}{JIT}{Just-In-Time}
\newacronym{lda}{LDA}{Linear Discriminant Analysis}
\newacronym{da}{DA}{Decoding Accuracy}
\newacronym{rpc}{RPC}{Remote Procedure Call}
\newacronym{brt}{BRT}{\texttt{beaverbuild}, \texttt{rysnc}, and \texttt{Titan}}
\newacronym{ep}{EP}{Exclusive Provider}
\newacronym{tee}{TEE}{Trusted Execution Environment}
\newacronym{il}{IL}{Inclusion List}
\newacronym{aps}{APS}{Attestor-Proposer Separation}
\newacronym{ea}{EA}{Execution Auction}
\newacronym{cr}{CR}{Censorship-Resistance}
\newacronym{hft}{HFT}{High-Frequency Trading}
\newacronym{l2}{L2}{Layer-2}
\newacronym{l1}{L1}{Layer-1}
\newacronym{bsc}{BSC}{Binance Smart Chain}
\newacronym{zk}{ZK}{Zero Knowledge}
\newacronym{mcp}{MCP}{Multiple Concurrent Proposers}
\newacronym{tvl}{TVL}{Total Value Locked}
\newacronym{sia}{SIA}{Sequence-Independent Arbitrage}
\newacronym{sda}{SDA}{Sequence-Dependent Arbitrage}
\newacronym{cdf}{CDF}{Cumulative Distribution Function}
\newacronym{ci}{CI}{Confidence Interval}
\newacronym{fp}{FP}{False Positive}
\newacronym{fn}{FN}{False Negative}
\newacronym{clob}{CLOB}{Central Limit Order Book}
\newacronym{p2p}{P2P}{Peer-to-Peer}
\newacronym{eip}{EIP}{Ethereum Improvement Proposal}
\newacronym{evm}{EVM}{Ethereum Virtual Machine}
\newacronym{abm}{ABM}{Agent-Based Modeling}
\newacronym{fdv}{FDV}{Fully Diluted Valuation}
\newacronym{cpmm}{CPMM}{Constant Product Market Maker}
\newacronym{fcfs}{FCFS}{First-Come, First-Served}
\newacronym{epbs}{ePBS}{enshrined Proposer-Builder Separation}
\newacronym{se}{SE}{Simulation Experiment}
\newacronym{exp}{EXP}{Experiment}
\newacronym{hhi}{HHI}{Herfindahl–Hirschman Index}
\newacronym{bal}{BAL}{Block-level Access List}
\newcommand{\encircled}[2][0.9mm]{%
    \raisebox{.8pt}{%
        \textcircled{%
            \raisebox{0.35pt}{%
                \kern #1
                \scalebox{0.70}{#2}
            }%
        }%
    }%
}
\makeatletter
\newcommand*{\ensquared}[1]{\relax\ifmmode\mathpalette\@ensquared@math{#1}\else\@ensquared{#1}\fi}
\newcommand*{\@ensquared@math}[2]{\@ensquared{$\m@th#1#2$}}
\newcommand*{\@ensquared}[1]{%
\tikz[baseline,anchor=base]{\node[draw,outer sep=0pt,inner sep=0.6mm,minimum width=3.8mm] {#1};}} 
\makeatother

\newcommand{\rE}{\mathrm{E}}
\newcommand{\rD}{\mathrm{D}}
\newcommand{\rS}{\mathrm{S}}
\newcommand{\Rset}{\mathbb{R}}
\newcommand{\cpsb}{\texttt{CPSB}}

\setcopyright{none}
\renewcommand\footnotetextcopyrightpermission[1]{}
\acmConference[WWW '27]{The ACM Web Conference 2027}{May 10--14, 2027}{Dublin, Ireland}
\acmBooktitle{The ACM Web Conference 2027}

\begin{document}

\title{More Meters or More Prices? One- vs. Multi-dimensional Fee Mechanisms on Ethereum}


\author{Fei Wu}
\authornote{Fei Wu performed work in part during an internship at Ethereum Foundation.}
\affiliation{%
  \institution{King's College London}
  \city{London}
  \country{UK}
}

\author{Maria Inês Silva}
\affiliation{%
  \institution{Ethereum Foundation}
  \city{Zurich}
  \country{Switzerland}
}

\author{Stefanos Leonardos}
\affiliation{%
  \institution{King's College London}
  \city{London}
  \country{UK}
}

\author{Carmine Ventre}
\affiliation{%
  \institution{King's College London}
  \city{London}
  \country{UK}
}



\begin{abstract}
Permissionless blockchains use transaction fee mechanisms to allocate the
resources that transactions consume, such as execution, data bandwidth, and
state creation. Recent blockchain designs explore pricing these resources separately rather than through a single unit. We study this shift on Ethereum, whose roadmap moves in stages from a one-dimensional fee mechanism to a multi-dimensional one. We build a demand model for the three resources, calibrated from Ethereum's 2025 block-capacity increases, and find that state creation is the most price-responsive. We then evaluate both mechanisms at higher block capacity, combining the model with transaction-level metering reconstruction and block-level simulation. Under a one-dimensional mechanism, state creation determines the fee and suppresses execution throughput. Adjusting how state is metered can align resource usage under constant demand, but a single fee cannot track all resources when demand varies, leaving all underutilized. A multi-dimensional mechanism removes this fee coupling, but execution still generates data that must be propagated. Across different allocations between block execution and propagation time, the multi-dimensional mechanism delivers 4--34\% more execution throughput at similarly low congestion, with higher target utilization and state growth near its intended level.
\end{abstract}


\maketitle

\section{Introduction}
\label{sec:intro}

Permissionless blockchains are shared infrastructure for decentralized Web applications run by a network of nodes, notably financial applications such as asset exchanges and lending markets. Users submit transactions to transfer assets or invoke \emph{smart contracts}, programs whose execution and state updates are validated by network participants. Transactions are processed in blocks and consume several resources. \emph{Execution} comprises computation and access to existing state. \emph{Data bandwidth} (\emph{data} for short) comprises bytes propagated with a block. \emph{State creation} (\emph{state} for short) comprises additions to persistent accounts, program code, and storage. 

Each block has limited capacity for these resources, and the transaction fee mechanism decides who uses it and at what price. The price adjusts toward a per-block \emph{target}, a sustainable average usage set below a hard \emph{limit}: the limit caps any single block and leaves room for bursts, while the target governs long-run load, such as how fast persistent state grows. Many chains use a \emph{one-dimensional} fee mechanism, which meters all resources in a common unit, such as \emph{gas}, and charges one price per unit toward one gas target. Yet the resources impose different costs: execution and data must be processed and propagated before each block's deadline, whereas state accumulates in storage that every node retains. And as block capacity scales and lower fees attract more activity, demand for each resource can grow differently, so one price may not keep each near its target. A growing number of blockchains are therefore exploring \emph{multi-dimensional} fee mechanisms, in which each resource has its own price and target~\cite{vitalik_post,eip7999,solana_chili_peppers,goren2024multidimensional}. This raises a natural question: as block capacity scales, how do one-dimensional and multi-dimensional fee mechanisms compare in execution throughput, resource utilization, and state-growth control?

\begin{figure}[t]
\centering
\definecolor{cE}{HTML}{4C72B0}\definecolor{cD}{HTML}{DD8452}\definecolor{cS}{HTML}{55A868}
\begin{tikzpicture}[font=\scriptsize, >=stealth,
  res/.style={draw=#1, thick, fill=#1!12, rounded corners=2pt, minimum width=1.45cm, minimum height=0.38cm, inner sep=1pt, font=\footnotesize},
  box/.style={draw, minimum width=1.55cm, minimum height=0.38cm, inner sep=1pt, align=center},
  fee/.style={draw, thick, fill=gray!12, minimum width=0.8cm, minimum height=0.38cm, inner sep=1pt, align=center},
  arr/.style={->, semithick},
  bal/.style={->, semithick, cD, dashed}]
\node[res=cE] (E) at (4.05,0) {Execution};
\node[res=cD] (D) at (4.05,-0.5) {Data};
\node[res=cS] (S) at (4.05,-1.0) {State};
\node[font=\footnotesize\bfseries] at (1.45,0.42) {(a) One-dimensional};
\node[box, minimum height=0.88cm] (R) at (2.1,-0.25) {Regular gas\\(data via floor)};
\node[box] (aSG) at (2.1,-1.0) {State gas};
\node[fee, minimum width=0.95cm, minimum height=1.38cm] (b) at (0.5,-0.5) {one\\fee $b$};
\draw[arr] (E.west) -- (E.west -| R.east);
\draw[arr] (D.west) -- (D.west -| R.east);
\draw[arr] (S.west) -- (aSG.east);
\draw[arr] (R.west) -- (R.west -| b.east);
\draw[arr] (aSG.west) -- (aSG.west -| b.east);
\node[font=\footnotesize\bfseries] at (7.05,0.42) {(b) Multi-dimensional};
\node[box] (EG) at (6.5,0) {Execution gas};
\node[box] (DG) at (6.5,-0.5) {Data gas};
\node[box] (SG) at (6.5,-1.0) {State gas};
\node[fee] (bE) at (7.975,0) {fee $b_E$};
\node[fee] (bD) at (7.975,-0.5) {fee $b_D$};
\node[fee] (bS) at (7.975,-1.0) {fee $b_S$};
\foreach \r/\m/\f in {E/EG/bE, D/DG/bD, S/SG/bS} {
  \draw[arr] (\r.east) -- (\m.west);
  \draw[arr] (\m.east) -- (\f.west);
}
\draw[bal] ([yshift=-3pt]E.east) -- node[pos=0.5, above, sloped, inner sep=0.5pt, font=\scriptsize]{BAL} ([yshift=3pt]DG.west);
\draw[bal] ([yshift=3pt]S.east) -- node[pos=0.5, below, sloped, inner sep=0.5pt, font=\scriptsize]{BAL} ([yshift=-3pt]DG.west);
\end{tikzpicture}
\caption{The three resources and how each mechanism meters and prices them. (a) Execution and data share one meter, and one base fee follows the more congested branch. (b) Each resource has its own meter and fee, but execution and state also generate block-level access list (BAL) data.}
\label{fig:concept}
\end{figure}
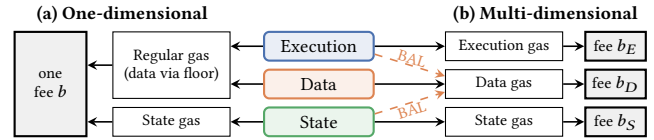

Ethereum is moving along this path in stages, which makes it a natural case
study. Under \acrshort{eip}-1559~\cite{eip1559}, a single base fee prices all gas and adjusts toward a per-block gas target, currently 30M against a 60M limit. The Glamsterdam upgrade raises the gas limit to 200M~\cite{glamsterdam}, with further increases planned~\cite{scalinghegota}, and meters state gas separately through \emph{state-gas metering} (\acrshort{eip}-8037~\cite{eip8037}). The fee nonetheless remains one-dimensional: one base fee applies to all resources and responds to whichever of state and the remaining gas is more congested (cf. \Cref{fig:concept}(a)). The proposed multi-dimensional fee mechanism (\acrshort{eip}-7999) gives execution, data, and state their own base fees and targets~\cite{eip7999pull} (cf. \Cref{fig:concept}(b)). 

As Ethereum scales block capacity toward its physical limits, the fee mechanism must allocate that capacity efficiently while keeping each resource's usage near its target and within its capacity. Existing analyses of these fee mechanisms have focused primarily on theoretical guarantees for welfare, capacity utilization, and price adjustment~\cite{angerismultidimensional,
lavee2025doesblockchainneedmultidimensional,onevsmultidimensional},
but do not establish how Ethereum's proposed mechanisms perform under observed resource demand and capacity constraints. We address this gap by providing the first empirical, simulation-based comparison of one- and multi-dimensional fee mechanisms, with resource demand calibrated from empirical data. 
\textbf{Contributions.} Our contributions are as follows.
\begin{itemize}[leftmargin=*, topsep=1pt, itemsep=1pt]
    \item \textbf{Demand calibration} (\Cref{sec:demand}). 
    Using Ethereum's gas-limit increases in 2025, which lowered fees for all resources at once, we estimate how demand for each resource responds to its price. Because these events changed all prices together, they cannot reveal substitution between resources, so we model each resource's demand independently. State is the most price-responsive resource and execution the least.

    \item \textbf{One-dimensional analysis} (\Cref{sec:one-dimensional}). 
    Under calibrated demand, as capacity grows and the fee falls, we find that state reaches the gas target first and sets the shared fee, leaving the other resources far below the target. Rescaling state metering lets all resources reach the gas target at the same fee when demand is constant, but when demand varies between blocks, one fee cannot track all resources, causing all resources to fall below target, even with higher targets or an alternative update rule.

    \item \textbf{Multi-dimensional analysis} (\Cref{sec:7999}). Separate fees remove this price coupling, but executing transactions generates data (BALs), so execution pays part of the data fee and is constrained by the data capacity. A high data fee caps how much execution any execution fee can attract, and when demand bursts fill the data limit, execution is excluded together with the data it generates. Allocating more of each block's time to data propagation relieves this until the shorter execution time binds.

    \item \textbf{Comparison} (\Cref{sec:comparison}). 
    At comparable congestion, the multi-dimensional mechanism delivers more execution while holding state growth near its target, which no one-dimensional configuration we test achieves alongside high execution.
\end{itemize}
More generally, under a one-dimensional mechanism, one fee cannot track all resources, causing all resources to be underutilized. Multi-dimensional fees remove this problem, but resources can remain coupled: when one resource generates demand for another, the first stays constrained by the second's price and capacity.

\noindent\textbf{Reproducibility.}
Code and data are available at
\href{https://github.com/M1kuW1ll/eip-7999-research}{\underline{link}}.



\section{Background and Related Work}
\label{sec:background}
Ethereum produces one block per 12-second \emph{slot}. A designated validator proposes and propagates a block, which the other nodes must receive and execute before deadlines within the slot. Transactions run in the \gls{evm}, where each operation consumes gas according to a protocol-defined schedule, and each block's gas usage is bounded by its gas limit. Protocol changes are specified as \glspl{eip}. \Cref{tab:eips} lists the upcoming changes considered in this paper. We distinguish \emph{metering}, which converts resource consumption into gas in one or more counters, from \emph{pricing}, which sets the fee charged per unit of gas. A one-dimensional mechanism can meter resources in separate counters while resources still share one price.

\begin{table}[t]
    \centering
    \footnotesize
    \setlength{\tabcolsep}{4pt}
    \begin{tabularx}{\columnwidth}{@{}ll>{\raggedright\arraybackslash}X@{}}
        \toprule
        EIP & Name used here & Role in this paper\\
        \midrule
        2780 & intrinsic gas repricing & execution metering\\
        7928 & \acrfullpl{bal} & execution- and state-generated data\\
        7999 & multi-dimensional fees & separate resource fees\\
        8037 & state-gas metering & separate state counter\\
        8038 & state-access gas repricing & execution metering\\
        8131 & static content floor & floor on static content\\
        8279 & \gls{bal} floor & \gls{bal} bytes in the floor\\
        8368 & \cpsb\ recalibration & growth-matched benchmark\\
        8372 & state-gas normalization & normalized benchmark\\
        \bottomrule
    \end{tabularx}
    \caption{Ethereum Improvement Proposals (EIPs) referenced in this paper, ordered by their EIP number.}
    \vspace{-0.4cm}
    \label{tab:eips}
\end{table}




\parhead{One-dimensional fee mechanism.} 
Under \gls{eip}-1559, all resource consumption contributes to a common gas counter, and transactions pay a minimum fee per gas, the \emph{base fee}, which adjusts toward a per-block gas target $T_G$, set to half the gas limit $L_G$~\cite{eip1559}. State-gas metering separates state gas from the remaining gas, which we call the \emph{regular branch}~\cite{eip8037}. State gas is metered by cost per state byte (\cpsb), with a reference value of 1,530 calibrated to 120 GiB annualized state growth under a 75M gas target.

Let $b_t$ denote the base fee, $g_{R,t}$ and $g_{S,t}$ the regular- and state-branch gas usage in block $t$, respectively. The mechanism retains a single base fee and the \gls{eip}-1559 update rule driven by the larger branch, $G_t=\max\{g_{R,t},g_{S,t}\}$. Ignoring integer rounding,
\begin{equation}
b_{t+1}
\approx
b_t\left(1+\frac{G_t-T_G}{8T_G}\right).
\label{eq:1559_update}
\end{equation}
The shared base fee increases when either branch exceeds the target and decreases only when both branches are below it.

\parhead{Multi-dimensional fee mechanism}
The multi-dimensional fee market was first proposed to price calldata separately~\cite{eip7999}. Its extension, building on state-gas metering, gives state, data, and execution distinct base fees~\cite{eip7999pull}. Execution and data have separate per-block limits, while state has only a target. Each base fee follows a normalized, exponentiated excess-gas counter. Let $b_{i,t}$ denote the base fee of resource $i$ at block $t$, $G_{i,t}$ its gas usage, and $T_i$ its target. Ignoring integer rounding, the base fee in the next block is
%
\begin{equation}
b_{i,t+1}
\approx
b_{i,t}
\exp\!\left[
2\ln(1.125)
\frac{G_{i,t}-T_i}{h_i}
\right],
\label{eq:7999_update}    
\end{equation} 
where the normalization denominator $h_i$ is the gas limit for data and execution and the target for state. The protocol sets the smallest fee unit of 1 wei = \(10^{-9}\) gwei = \(10^{-18}\) ETH.

\parhead{Block capacity}
Data propagation and execution processing must fit within block-validation deadlines. Within the shared 9-second budget in the Glamsterdam specification~\cite{eip7732}, increasing propagation time $\tau$ raises data capacity but reduces execution capacity. The propagation model calibrated on mainnet blocks gives a data capacity per block of $B_{\max}(\tau)=1024\,\frac{1000\tau-569}{0.443}$~bytes~\cite{propagation_model}, and an execution throughput of 100M gas per second gives an execution capacity of $L_\rE(\tau)=100(9-\tau)$M gas~\cite{eip7904}. We restrict the analysis to $\tau\leq5$ seconds, where $B_{\max}=9.77$~MiB approaches the 10~MiB consensus-layer limit~\cite{consensus_p2p}.

\parhead{Data pricing}
The Glamsterdam upgrade introduces block-level access lists (\glsunset{bal}BALs, \gls{eip}-7928), which record each block's accessed accounts, storage, and state changes to support parallel execution~\cite{eip7928}. Execution and state creation produce BALs, which must propagate with the block like other data.\footnote{We use the transaction-level runtime \gls{bal} meter, which enters the transaction floor~\cite{eip8279}; the encoded block-level object may differ slightly in size.}
 

Under the one-dimensional mechanism, data has no separate price. Instead, it enters the \emph{transaction floor}~\cite{eip7623}: a transaction is charged the larger of its execution gas and a floor equal to its intrinsic gas, the fixed base cost of every transaction, plus $F$ gas per byte of static content, user-submitted bytes such as calldata and access lists (content floor, \gls{eip}-8131~\cite{eip8131}), and of runtime \gls{bal} (\gls{bal} floor, \gls{eip}-8279~\cite{eip8279}). We call $F$ the floor rate. A transaction is \emph{data-exposed} when the floor is larger. Otherwise, adding data bytes leaves its gas unchanged until the floor overtakes its execution gas.

The baseline rate is $F=64$ gas per byte~\cite{eip8131,eip8279}. Since a data-exposed transaction pays at least $F$ gas per byte, the floor bounds the gas limit by $L_G\leq FB_{\max}(\tau)$, and raising $F$ admits a higher gas limit. The gain stops near $F=96$: an ETH transfer uses 21,000 gas for 221 bytes~\cite{scalinghegota}, so beyond that rate a block of transfers sets the constraint. The largest feasible gas limit and the smallest integer rate supporting it are
\begin{equation}
    L_G^\ast(\tau)=\min\left\{L_\rE(\tau),\,\frac{21{,}000}{221}B_{\max}(\tau)\right\},
    \quad
    F^\ast(\tau)=\left\lceil\frac{L_G^\ast(\tau)}{B_{\max}(\tau)}\right\rceil.
    \label{eq:floor-selection}
\end{equation}

Under the multi-dimensional fee mechanism, data and execution have their own base fee, target, and limit. 
In \gls{eip}-7999, all data components consume 16 data gas per byte \cite{eip7999pull}. 
We apply the same rate to runtime-metered \gls{bal} bytes.
State has no per-block limit, but its usage is controlled with its base fee and target in the long term \cite{eip7999pull}.
\Cref{tab:shared-capacity} reports the resulting limits under each mechanism.

\begin{table}[t]
    \centering
    \footnotesize
    \setlength{\tabcolsep}{4pt}
    \begin{tabular}{@{}ccrrrrr@{}}
        \toprule
        & & $F=64$
        & \multicolumn{2}{c}{Adjusted-floor}
        & \multicolumn{2}{c}{Multi-dimensional} \\
        \cmidrule(lr){3-3}
        \cmidrule(lr){4-5}
        \cmidrule(l){6-7}
        $\tau$ (s) & $B_{\max}$ (MiB)
        & $L_G$ & $F^\ast$ & $L_G^\ast$
        & $L_E$ & $L_D$ \\
        \midrule
        3.0 & 5.36 & 359.6M & 96 & 534.0M & 600.0M & 90.0M  \\
        3.5 & 6.46 & 433.6M & 82 & 550.0M & 550.0M & 108.4M \\
        4.0 & 7.56 & 500.0M & 64 & 500.0M & 500.0M & 126.9M \\
        4.5 & 8.67 & 450.0M & 50 & 450.0M & 450.0M & 145.4M \\
        5.0 & 9.77 & 400.0M & 40 & 400.0M & 400.0M & 163.9M \\
        \bottomrule
    \end{tabular}
    \caption{Data capacity and gas limits under each propagation
    allocation. One-dimensional benchmarks use a common gas
    limit under specified $F$, while the multi-dimensional mechanism uses separate
    execution and data limits, $L_E$ and $L_D$.}
    \label{tab:shared-capacity}
    \vspace{-0.4cm}
\end{table}

\parhead{A toy example}
Let execution and state each have a target of 2 gas per block, with demands of $p^{-1/4}$ and $p^{-1/2}$ units at price $p$ per unit, so state is more price-responsive. State is also metered more heavily, at 4 gas per unit against 1. Under a shared fee $b$, state uses $4(4b)^{-1/2}=2b^{-1/2}$ gas and fills its target at $b=1$, whereas execution uses $b^{-1/4}$ gas and would fill its target only at $b=1/16$. The fee settles at $b=1$, leaving execution at half its target. Separate fees settle at $b_S=1$ and $b_E=1/16$, filling both targets. Now let each unit of execution generate $1/4$ gas of \gls{bal} data, and let static-content demand hold the data fee at 1. Execution then pays at least $1/4$ per unit even at a zero execution fee, reaching at most $4^{1/4}\approx1.41$ gas, $71\%$ of its target. When the data limit binds, excess \gls{bal} can be dropped only with the execution that generates it. Separate fees decouple the prices, not the resources.

\parhead{Related work}
Prior work studies  \gls{eip}-1559~\cite{roughgarden2024transactionfees,1559honeymoon,1559dynamicalanalysis,optimaldespitechaos1559,Liu_empirical1559_2022}, resource usage and gas
metering on Ethereum ~\cite{maria_ethreserach_empirical,heimbach2026evmworkloadswildevidence}, and resource demand elasticities from within-wallet fee variation on Arbitrum~\cite{anchuri2026elasticity}. Theoretical work on multi-dimensional
fees derives resource prices, bounds their welfare, and quantifies the capacity lost by combining resource constraints~\cite{diamandis2023,angerismultidimensional,lavee2025doesblockchainneedmultidimensional,onevsmultidimensional}. We evaluate the resulting utilization trade-offs on Ethereum using empirically calibrated demand and simulation: a one-dimensional fee mechanism forgoes execution even when separate meters align its target-clearing fees, and a multi-dimensional fee mechanism recovers much of it.

\section{Demand Model and Calibration}
\label{sec:demand}

We develop a demand model for state, static content data, and execution, denoted by $\rS$, $\rD$, and $\rE$, with $\Rset = \{\rS, \rD, \rE\}$. Let $q_i$ denote resource $i$'s activity in \emph{historical gas-equivalent units}, the gas it consumed under the historical gas metering, so that the three components sum to block gas usage and activity is comparable across metering rules. Let $p_i$ denote the effective price of one unit, the base fee charged on the gas it is metered at, and let $p^0$ and $q_i^0$ denote the historical fee and per-block activity around which we calibrate.

Demand for each resource can depend on all three prices, $q_i = D_i(p_\rS, p_\rD, p_\rE)$, with \emph{own-price elasticities} $\epsilon_{ii}$ and \emph{cross-price elasticities} $\epsilon_{ij}$, $i\neq j$, where $\epsilon_{ij}\equiv-\partial\ln D_i/\partial\ln p_j$. Cross-price elasticities are negative for substitutes and positive for complements.
Under the current Ethereum one-dimensional fee market, all resources face the same base fee $b$, 
so a block capacity increase moves their effective prices by the same proportion. The observed response to a common base fee change is
\begin{equation}
  \epsilon_i^{\mathrm{common}}
  \;\equiv\;
  -\,\frac{d \ln D_i(b,b,b)}{d \ln b}
  \;=\;
  \sum_{j \in \Rset} \epsilon_{ij},
  \label{eq:row-sum}
\end{equation} 
the sum of the own- and cross-price elasticities. 
We adopt an \emph{independent demand model}, following the prior analysis of Ethereum state demand~\cite{silva2026stateburst}.

\begin{assumption}[Independent demand]
\label{ass:independence}
Each resource's demand depends only on its own effective price:
$\epsilon_{ij}=0$ for $i\neq j$, so that
$\epsilon_i\equiv\epsilon_{ii}=\epsilon_i^{\mathrm{common}}$.
\end{assumption}

As Ethereum never changed relative resource prices, the common-fee variation cannot separate own- from cross-price effects.\footnote{The calldata floor~\cite{eip7623} changed the data price only minimally for 2\% of transactions.} Therefore, independence is a modeling assumption: it does not claim that users demand resources separately, since a transaction can consume all three. We model each resource using a constant-elasticity demand curve~\cite{varian2024microeconomics}:
\begin{equation}
q_i(p_i)=q_i^0\left(\frac{p_i}{p^0}\right)^{-\epsilon_i},
\qquad p_i>0,\quad\epsilon_i>0.
\label{eq:demand}
\end{equation}
A one-percent fall in $p_i$ raises $q_i$ by about $\epsilon_i$ percent, and we hold $\epsilon_i$ constant over the prices considered. When a common fee falls, resources with larger elasticities grow faster, so additional capacity changes the composition of activity as well as its amount.


\parhead{Elasticity calibration}
We use Ethereum mainnet data on resource activity, gas usage, and base fees from Xatu~\cite{xatu} around the three block-capacity increases of 2025: 30M to 36M on February 4, 36M to 45M on July 21, and 45M to 60M on November 25. We exclude the November event, whose post-period overlaps a blob-capacity increase~\cite{bpo1} that can shift calldata demand independently. For each retained event, we compare 35-day windows before and after the increase, excluding the event day and three days on each side. Within each window, we measure activity as total resource gas per block and the fee as the median of daily median base fees, capturing adjustment over weeks rather than block-to-block fluctuation.

\Cref{tab:event-elasticities} reports each event's responses of resource activity, $r_{i,e}$, and of total gas, $a_e$, the log change in activity per log fall in the fee. Since the two events measure the same constant elasticities, we average their resource responses and add 0.004 to each, so that the model's elasticity of total gas, which averages the three by their shares of gas, matches the observed total-gas response (cf. \Cref{appendix:elasticity_calibration}). This gives $\epsilon_\rE=0.121$, $\epsilon_\rD=0.229$, and $\epsilon_\rS=0.335$. Execution exhibits the largest difference between the two event responses, 0.011 and 0.223. The data elasticity applies to static content, since \glspl{bal} are not observed historically. Across 21-, 60-, and 75-day windows, state remains the most price-responsive and execution the least (cf. \Cref{appendix:elasticity_sensitivity}). The ordering breaks only in the excluded November event, where data falls below execution, as expected if cheaper blob space drew rollup data away from calldata.

\begin{table}[t]
  \centering
  \footnotesize
  \setlength{\tabcolsep}{4pt}
  \begin{tabular}{@{}lrrrrr@{}}
    \toprule
    Event
      & $\ln(p_e^{+}/p_e^{-})$
      & $r_{\rE,e}$
      & $r_{\rD,e}$
      & $r_{\rS,e}$
      & $a_e$ \\
    \midrule
    30M $\to$ 36M, Feb.~4 & $-2.114$ & 0.011 & 0.128 & 0.322 & 0.087 \\
    36M $\to$ 45M, Jul.~21 & $-0.887$ & 0.223 & 0.323 & 0.339 & 0.251 \\
    Calibrated $\widehat{\epsilon}_i$ & & 0.121 & 0.229 & 0.335 & \\
    \midrule
    45M $\to$ 60M, Nov.~25 & $-1.254$ & 0.180 & 0.086 & 0.413 & 0.234 \\
    \bottomrule
  \end{tabular}
    \caption{Event responses using nominal 35-day windows, excluding the event day and three days on each side. The third event is excluded from calibration.}
  \label{tab:event-elasticities}
  \vspace{-0.4cm}
\end{table}

\parhead{Mechanism-specific metering and prices.}
Proposed repricings (cf. \Cref{tab:eips}) change the gas that an identical activity consumes. Measured in historical gas-equivalent units, activity is comparable across gas metering rules: a metering coefficient $m_i$ converts it into counterfactual metered gas $g_i = m_iq_i$, so a unit of activity metered at $m_i$ gas and charged $b_i$ per gas has effective price $p_i=m_ib_i$. We calibrate these coefficients when introducing each mechanism.

\section{One-dimensional Fee Mechanisms}
\label{sec:one-dimensional}



We study three one-dimensional benchmarks, each with a single base fee updated by the larger of the regular and state branches (cf. Eq.~\eqref{eq:1559_update}), under the capacities in \Cref{tab:shared-capacity}. The \emph{baseline} retains the specified floor rate $F=64$ and $\cpsb=1530$~\cite{eip8037}. The other two use the adjusted floor rates $F^\ast$. The \emph{growth-matched} benchmark recalibrates \cpsb\ so that state usage at the target corresponds to 120~GiB/year, as in \cpsb\ recalibration (\acrshort{eip}-8368~\cite{eip8368}). The \emph{normalized} benchmark, following state-gas normalization (\acrshort{eip}-8372~\cite{eip8372}), further scales \cpsb\ and the state-gas limit by a common factor chosen from calibrated demand.

\subsection{Metering and calibration}
\label{sec:shared-metering}

Under the shared base fee $b$, each activity faces effective price $m_ib$, so its metered gas is
\begin{equation}
    m_i\,q_i(m_ib)=q_i^0\,m_i^{\,1-\epsilon_i}\left(\frac{b}{p^0}\right)^{-\epsilon_i},
    \qquad i\in\Rset.
    \label{eq:metered}
\end{equation}
We calibrate historical price and activity anchors $p^0$ and $q_i^0$ on mainnet blocks from 1 February to 31 May 2026, covering 120 days and 860,505 blocks: $p^0=0.1069$~gwei is the median of daily median base fees, and the per-block activity anchors are 5.24M, 1.18M, and 23.94M historical gas-equivalent units for state, data, and execution.


\parhead{Execution and state}
Replaying these blocks under the state-access (\acrshort{eip}-8038) and intrinsic-gas (\acrshort{eip}-2780) repricings gives $m_\rE=1.448$~\cite{eip8038,eip2780}, and state-gas metering gives
$m_\rS=5.656$ at $\cpsb=1530$~\cite{eip8037}. Since state gas is proportional to \cpsb, $m_\rS(c)=\frac{5.656\,c}{1530}$. We detail transaction-level reconstruction in \Cref{appendix:metering_calibration}.

\parhead{Floor-metered data}

Because the floor is a maximum evaluated per transaction, the data contribution cannot be obtained by applying $F$ to aggregate bytes. Therefore, we reconstruct execution gas, static content bytes, and runtime \gls{bal} bytes for each transaction, which requires complete access and authorization lists~\cite{eip8131} and state traces~\cite{eip8279} from an Erigon archive node. Since tracing is computationally expensive, we randomly sample 50 blocks per day from the anchor period, covering 6,000 blocks and 1,899,748 transactions. Define $m_\rD^{\mathrm{floor}}(F)$ as the resulting mean data-attributed regular gas divided by $q_\rD^0$. The coefficient rises with the floor rate, from 1.589 at $F=40$ to 2.162 at $F=64$ and 4.033 at $F=96$ (\Cref{appendix:data_metering}). The regular branch metered gas is then
\begin{equation}
    R(b) = m_\rE\, q_\rE(m_\rE b)
         + m_\rD^{\mathrm{floor}}(F)\, q_\rD\!\bigl(m_\rD^{\mathrm{floor}}(F)\, b\bigr).
    \label{eq:regular-branch}
\end{equation}
We fold the \gls{bal}-related floor uplift into $m_\rD^{\mathrm{floor}}(F)$, assuming it follows static-data activity: at $F=96$, runtime \gls{bal} changes regular gas for $6.79\%$ of sampled transactions and raises aggregate regular gas by $2.75\%$, mostly in transactions already at or near the static floor. We detail data metering steps and BAL-related floor exposure in \Cref{appendix:data_metering}. The multi-dimensional model includes separately charged \gls{bal} in the effective prices of its parent activities.



\subsection{Equilibrium under a shared base fee}
\label{subsec:shared-equilibrium}

Let $S(b)=m_\rS q_\rS(m_\rS b)$ and $R(b)$ from Eq.~\eqref{eq:regular-branch} denote state and regular demand in metered gas. Both are continuous and strictly decreasing in $b$. Define their unique branch-clearing fees by $R(b_R)=T_G$ and $S(b_S)=T_G$. From Eq.~\eqref{eq:metered}, 
\begin{equation}     
b_S=p^0 m_\rS^{\frac{1-\epsilon_\rS}{\epsilon_\rS}}         
\left(\frac{q_\rS^0}{T_G}\right)^{1/\epsilon_\rS},     
\label{eq:state-clearing-fee} 
\end{equation} 
while $b_R$ is solved numerically. Under fixed demand and ignoring integer rounding, the unique \emph{fixed-demand equilibrium fee} is $b^\ast=\max\{b_R,b_S\}$, where $\max\{R(b^\ast),S(b^\ast)\}=T_G$.

Under the central elasticity calibration, state determines the equilibrium fee in the baseline and growth-matched benchmarks across all five capacity allocations. At the regular-clearing fee, $S(b_R)/T_G$ ranges from $5.4$ to $8.4$ and $13.4$ to $19.1$, respectively, so the fee must rise well above $b_R$ to clear state demand, leaving regular gas below target. Two features of the calibration produce this gap. State has the largest metering coefficient, and Eq.~\eqref{eq:state-clearing-fee} gives $\partial\ln b_S/\partial\ln m_\rS=(1-\epsilon_\rS)/\epsilon_\rS\approx1.99$, so heavier state metering raises its clearing fee more than proportionally. State is also the most price-responsive resource, so its activity expands fastest as the fee falls. State's control of the shared base fee holds under every elasticity calibration (cf. \Cref{appendix:elasticity_sensitivity}).


The two benchmarks differ in state growth. When state sets the fee, state creation equals $T_G/\cpsb$ bytes per block, so retaining $\cpsb=1530$ at a larger target lets the baseline grow state by 2.4 to 3.4 times its 120~GiB/year objective. The growth-matched benchmark restores the objective by raising \cpsb, but by the elasticity above, this raises $b_S$ and further suppresses regular gas.


The shared fee also transmits state demand to execution. If state demand shifts by a factor $a>0$ while state remains the bottleneck,
\begin{equation}
    b^\ast(a)=b_S a^{1/\epsilon_\rS},
    \qquad
    \frac{q_\rE^\ast(a)}{q_\rE^\ast(1)}=a^{-\epsilon_\rE/\epsilon_\rS},
    \label{eq:shared-spillover}
\end{equation}
so a sustained $10\%$ rise in state demand reduces equilibrium execution by about $3.4\%$, although execution demand is unchanged. Adjusting the floor rate cannot relieve this: at a fixed target, $b_S$ does not depend on $F$. Larger gas limits under adjusted floor leave state the bottleneck at every allocation, and the regular branch consequently remains below target despite having unused capacity.


\parhead{Normalizing state gas}
\label{subsec:eip8372-calibration}
The normalized benchmark addresses this mismatch by giving state gas a separately calibrated higher limit \cite{eip8372}. To maintain the physical state growth target, \cpsb\ is increased proportionally to the state gas limit. Starting from the growth-matched coefficient $c_0$, for a scale $k>1$ set
\begin{equation}
    L_S^{\mathrm{raw}}=kL_G,
    \qquad c=kc_0,
    \qquad \frac{L_S^{\mathrm{raw}}}{c}=\frac{L_G}{c_0},
    \label{eq:eip8372-scaling}
\end{equation}
where $L_S^{\mathrm{raw}}$ is the calibrated raw state-gas limit before normalization. The maximum state bytes per block, $L_S^{\mathrm{raw}}/c=L_G/c_0$, and the state bytes at target are unchanged, but state pays the effective price $km_{\rS,0}b$ while entering the fee update as $N_S(b;k)=m_{\rS,0}\,q_\rS(km_{\rS,0}b)$. Thus, the normalized state gas clears at
$b_S/k$. Choosing
\begin{equation}
    k^\ast=\frac{b_S}{b_R}
    \quad\Longrightarrow\quad
    R(b_R)=N_S(b_R;k^\ast)=T_G
    \label{eq:eip8372-alignment}
\end{equation}
aligns both branches at fee $b_R$. The state-byte price is unchanged in equilibrium, since $k^\ast c_0b_R=c_0b_S$, so state growth stays at its annualized target while the lower fee lets regular activity rise to the gas target. 
We report the full calibration of one-dimensional benchmarks for all capacity allocations in \Cref{appendix:one_dimensional}, \Cref{tab:one-dimensional-calibration}.

\subsection{Simulation under fluctuating demand}
In practice, resource demand varies between blocks. To evaluate the mechanisms' long-term performance when demand varies, we simulate all three one-dimensional benchmarks on the same block-level demand paths, constructed from mainnet activity in April--May 2026. Each run has a one-day burn-in for the fee to settle near the equilibrium, followed by seven measured days of 50,400 blocks, repeated over 32 bootstrap paths under the central elasticity calibration. We report average outcomes across the measured blocks and paths. The demand path construction is detailed in \Cref{appendix:demand_paths}. 


\begin{table}[t]
    \centering
    \footnotesize
    \setlength{\tabcolsep}{3pt}
    \renewcommand{\arraystretch}{0.95}
    \resizebox{\columnwidth}{!}{
    \begin{tabular}{@{}llcccccc@{}}
    \toprule
    Mechanism & $\tau$ (s)
        & \shortstack{Limit\\(M gas)}
        & \shortstack{Equilibrium\\fee (wei)}
        & \shortstack{Execution\\(M gas)}
        & \shortstack{Data\\(M gas)}
        & \shortstack{State\\(M gas)}
        & \shortstack{State growth\\(GiB/year)}\\
    \midrule
    \multirow{5}{*}{Baseline}
        & 3.0 & 359.6 & 86,981 & 82.0 & 12.0 & 180.1 & 288.1\\
        & 3.5 & 433.6 & 49,756 & 87.9 & 13.7 & 218.2 & 349.0\\
        & 4.0 & 500.0 & 32,513 & 92.6 & 15.1 & 252.2 & 403.4\\
        & 4.5 & 450.0 & 44,535 & 89.1 & 14.0 & 226.6 & 362.5\\
        & 5.0 & 400.0 & 63,308 & 85.3 & 12.9 & 200.9 & 321.4\\
    \midrule
    \multirow{5}{*}{\shortstack[l]{Growth-\\matched}}
        & 3.0 & 534.0 & 332,742 & 70.0 & 14.4 & 270.5 & 121.6\\
        & 3.5 & 550.0 & 323,038 & 70.2 & 11.8 & 278.8 & 121.6\\
        & 4.0 & 500.0 & 355,342 & 69.4 & 8.8 & 253.3 & 121.6\\
        & 4.5 & 450.0 & 394,824 & 68.5 & 7.3 & 227.9 & 121.5\\
        & 5.0 & 400.0 & 444,177 & 67.5 & 6.6 & 202.3 & 121.4\\
    \midrule
    \multirow{5}{*}{Normalized}
        & 3.0 & 534.0 & 88.22 & 166.9 & 71.2 & 196.1 & 88.1\\
        & 3.5 & 550.0 & 48.37 & 177.9 & 65.5 & 197.0 & 86.0\\
        & 4.0 & 500.0 & 56.37 & 174.7 & 48.3 & 179.2 & 86.0\\
        & 4.5 & 450.0 & 92.21 & 165.3 & 37.3 & 163.2 & 87.0\\
        & 5.0 & 400.0 & 190.52 & 151.9 & 29.4 & 146.4 & 87.8\\
    \bottomrule
    \end{tabular}}
    \caption{Equilibrium fees and simulated outcomes of the three one-dimensional benchmarks, with a target ratio of 1/2.
    State gas is normalized in the normalized benchmark.}
    \label{tab:one-dim-outcomes}
    \vspace{-0.4cm}
\end{table}

\Cref{tab:one-dim-outcomes} summarizes the outcomes of all three benchmarks. 
In the baseline and growth-matched benchmarks, as predicted in the equilibrium analysis, state is the bottleneck in more than $90\%$ of simulated blocks, and the regular branch is underused. The baseline grows state by 288.1--403.4~GiB/year. The growth-matched benchmark holds state growth near 120~GiB/year, with a small overshoot from the update rule~\cite{optimaldespitechaos1559}, but cuts execution to 67.5--70.2M. 

Normalizing state gas substantially increases execution activity by setting the regular branch clearing fee as the equilibrium fee. Execution gas reaches 177.9M at 3.5-second allocation, against 87.9M in the baseline and 70.2M in the growth-matched benchmark. Yet both branches remain below target on average: regular utilization is $88.5\%$--$90.6\%$ and normalized state utilization is $71.6\%$--$73.4\%$, giving state growth of 86.0--88.1~GiB/year.


The underutilization reflects resource price coupling in the one-dimensional mechanism through a single fee. When either branch exceeds the target, the fee rises for both, and the mechanism cannot lower the price of the underutilized branch independently. Normalization aligns the target-clearing fees under fixed demand, but does not preserve this alignment as demand varies. Consequently, both branches fall below target on average. State exhibits a higher level of underutilization because, as the most price-responsive resource, it contracts more when regular activity pushes the fee up.


\parhead{Varying the target-to-limit ratio}
\label{appendix:target-ratio}
Raising the ratio from 1/2 to 4/5 increases equilibrium execution by only $18.5\%$ in the baseline and $6\%$ in the growth-matched benchmark, since state still controls the fee, so we vary it only for the normalized benchmark, recalibrating \cpsb\ and $k$ at each new gas target. A higher ratio lowers the maximum state bytes per block relative to the target, from twice the target quantity at 1/2 to 1.25 times at 4/5, and lowers the largest upward fee step from $12.5\%$ to $3.125\%$.

\begin{figure}[t]
    \centering
    \includegraphics[width=\linewidth]{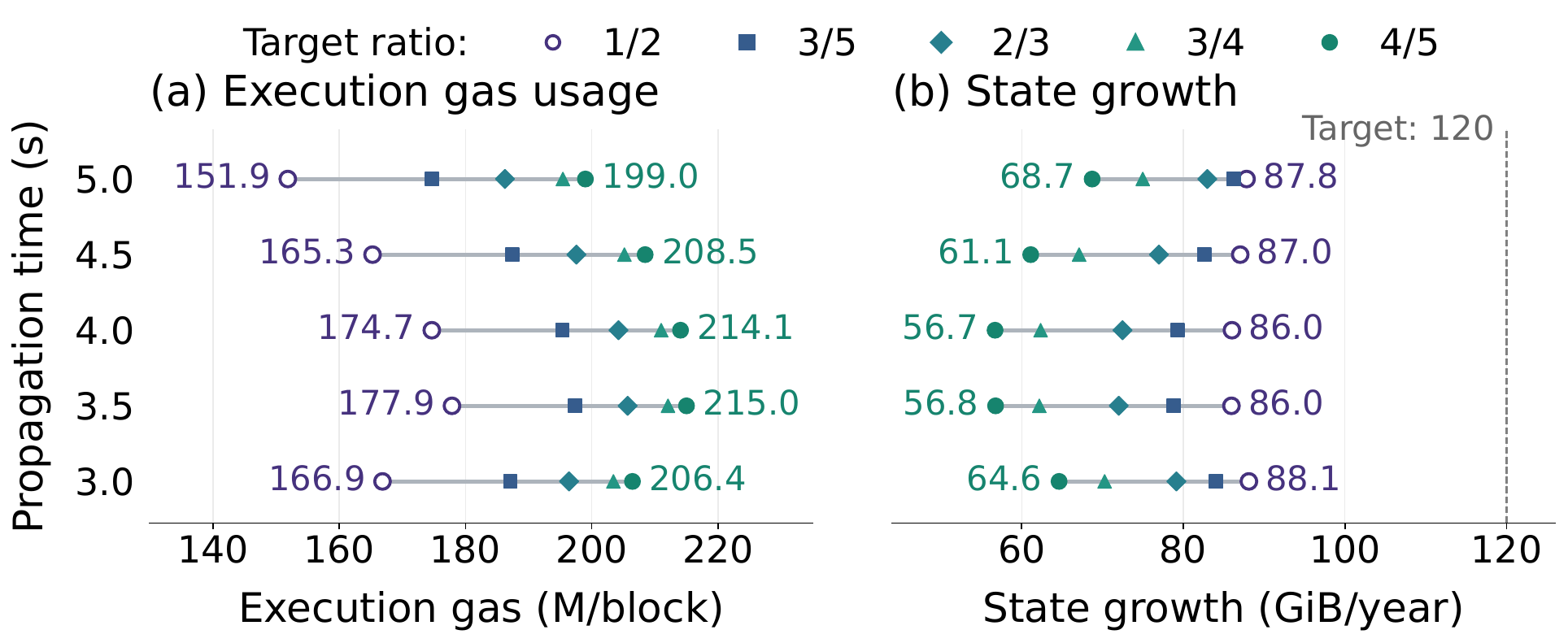}
    \caption{Normalized benchmark across target ratios and
    propagation allocations. (a) Mean execution gas. (b) Annualized state growth, with the dashed line marking the target.}
    \label{fig:8372-ratio}
\end{figure}

In \Cref{fig:8372-ratio}, execution rises with the target ratio at every allocation: at 3.5 seconds, it reaches 215.0M with a ratio of 4/5, against 177.9M with a ratio of 1/2, while state growth falls from 86.0 to 56.8~GiB/year. Congestion rises as well: with ratios of 2/3 and above, more than $10\%$ of blocks reach a hard limit, against about $4\%$ with a ratio of 1/2. With a ratio of 4/5, the integer update also holds the fee above its continuous equilibrium of 2.41~wei: a downward step needs a shortfall of at least $8/b$ of the target, so downward adjustments round to zero once the fee is below 8 wei. The higher fee deepens the state shortfall, since state contracts most when the shared fee is high, as does the lower maximum of state bytes per block, which clips positive demand shocks. \Cref{sec:comparison} reports results under an alternative update rule, detailed in \Cref{appendix:update-rule}.

\section{Multi-dimensional Fee Mechanism}
\label{sec:7999}
Unlike preceding one-dimensional benchmarks, the multi-resource extension of \gls{eip}-7999 gives execution, data, and state separate base fees~\cite{eip7999pull}. State keeps the metering of \gls{eip}-8037, with $\cpsb=1530$ and $m_S=5.656$, and a target of $T_S=75$M gas per block, corresponding to 120~GiB of annualized state growth, but no per-block limit. Execution and data limits follow the propagation allocation (cf. \Cref{tab:shared-capacity}), and their targets $T_E<L_E(\tau)$ and $T_D<L_D(\tau)$ are design choices, so their target-to-limit ratios need not be 1/2. Separate fees do not make the resources independent: execution and state activities generate \gls{bal} data, so the price and capacity of data still constrain them.

\subsection{Demand with \gls{bal} charge}

Under the multi-dimensional fee mechanism, \gls{bal} is directly metered in the data dimension and charged the data base fee. Unlike static content, \gls{bal} is not demanded by users but generated by the protocol from transaction activity: users can change how much execution or state creation to undertake, but cannot retain the activity and discard the \gls{bal} it produces. Therefore, we include its charge in the price of the parent activity. 

Assuming that state-access-related \gls{bal} scales with execution and state-creation-related \gls{bal} with state, let $w_E$ and $w_S$ denote the \gls{bal} data gas generated per historical unit of execution and state. Reconstructing runtime \gls{bal} on the same 6,000 historical blocks gives 1.92M data gas per block at the anchor, of which 88.6\% is generated by execution and 11.4\% by state creation, so $w_E=0.0710$ and $w_S=0.0417$. With resource base fees $b_E$, $b_D$, and $b_S$, the effective resource prices are
\begin{equation}
    p_E = m_E b_E+w_E b_D,
    \quad
    p_S = m_S b_S+w_S b_D,
    \quad
    p_D = m_D b_D,
    \label{eq:multidim-effective-prices}
\end{equation}
where $m_D=1.807$ meters only static content data and differs from $m_D^{\mathrm{floor}}(F)$ in the one-dimensional model. We retain the demand curves of Eq.~\eqref{eq:demand}, assuming that the same responses apply to these \gls{bal}-inclusive prices. The construction of $m_D$ and decomposition of runtime \gls{bal} are detailed in \Cref{appendix:data_metering}.

Metered resource demand is $g_E = m_E q_E$ for execution, $g_S=m_S q_S$ for state, and $g_D = m_D q_D+w_E q_E+w_S q_S$ for data. The resource coupling runs both ways: more execution generates more data demand, while a higher data fee raises the price of execution and reduces its demand. The same holds for state creation. Moreover, a transaction cannot be included without the \gls{bal} it generates. When the data limit binds, \emph{bundle inclusion} therefore removes execution and state activity together with their \gls{bal}, even if the execution limit has not been reached.

\subsection{Execution--data target frontier}
\label{subsec:multidim-equilibrium}
Under constant demand, a fee above the 1-wei minimum keeps adjusting until its resource reaches the target. At 1 wei, the fee cannot fall further and usage may remain below target. Ignoring integer rounding, the equilibrium conditions are
\begin{equation}
        b_i \geq b_{\min},
        \quad 
        g_i\leq T_i, 
        \quad
        (b_i-b_{\min})(T_i-g_i)=0,
        \quad 
        i\in\{E,D,S\},
    \label{eq:multidim-equilibrium}
\end{equation}
with $b_{\min}=1$~wei per gas. We focus on execution because the scaling configurations raise execution targets while retaining the 75M state target, and execution is the least price-responsive.

To determine whether all three targets can be filled, suppose that execution and state reach their targets. They generate $g_{\mathrm{BAL}}^\ast=(w_E/m_E)T_E+(w_S/m_S)T_S$ of
\gls{bal} data gas. The data fee clearing the remaining static content and the candidate fees of
execution and state, $i\in\{E,S\}$, are then
\[
    \widehat b_D=\frac{p^0}{m_D}
    \left(\frac{m_Dq_D^0}{T_D-g_{\mathrm{BAL}}^\ast}\right)^{1/\epsilon_D},
    \quad
    \widehat b_i=\frac{1}{m_i}\left[p^0\left(\frac{m_iq_i^0}{T_i}\right)^{1/\epsilon_i}
    -w_i\widehat b_D\right].
\]
All targets clear if and only if $T_D>g_{\mathrm{BAL}}^\ast$ and each candidate fee is at least $b_{\min}=1$~wei.


\begin{figure}[t]
    \centering
    \includegraphics[width=1\linewidth]{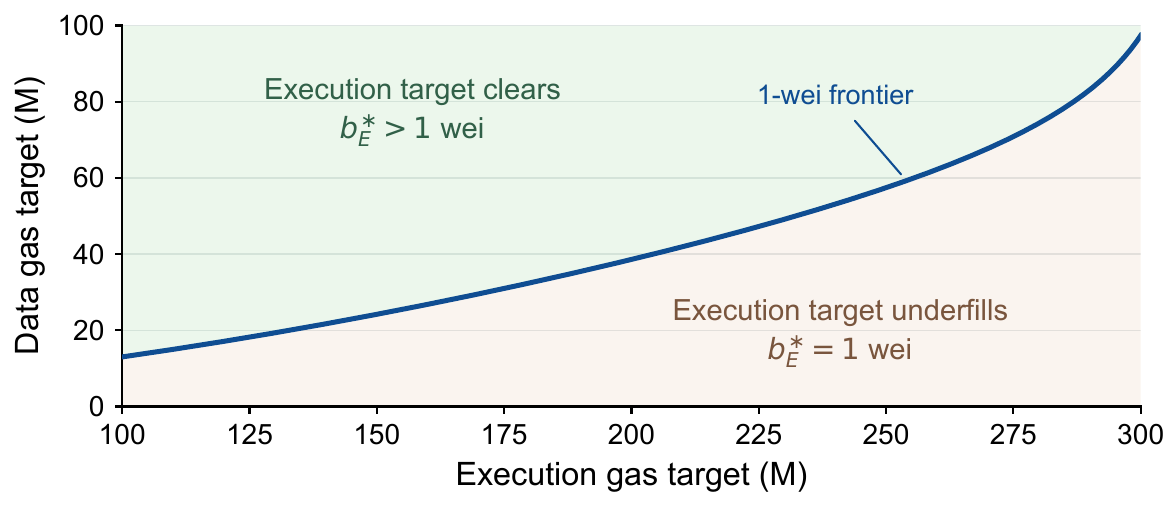}
    \caption{Execution--data target frontier of \gls{eip}-7999 under the central demand
    calibration and a 75M state target.}
    \label{fig:7999-frontier}
\end{figure}

At a fixed data target, a higher execution target lowers $\widehat b_E$ in two ways: it lowers the effective price needed to attract more execution, and the additional \gls{bal} leaves less room for static content, raising $\widehat b_D$. A higher data target does the reverse. The boundary
$\widehat b_E=b_{\min}$, shown in \Cref{fig:7999-frontier}, therefore gives the largest execution target that clears at each data target, provided the data and state fees remain feasible. Beyond this boundary, the execution fee sits at the 1-wei minimum and execution stays below target even in equilibrium.

\begin{figure*}[!t]
\centering
\includegraphics[width=1\linewidth]{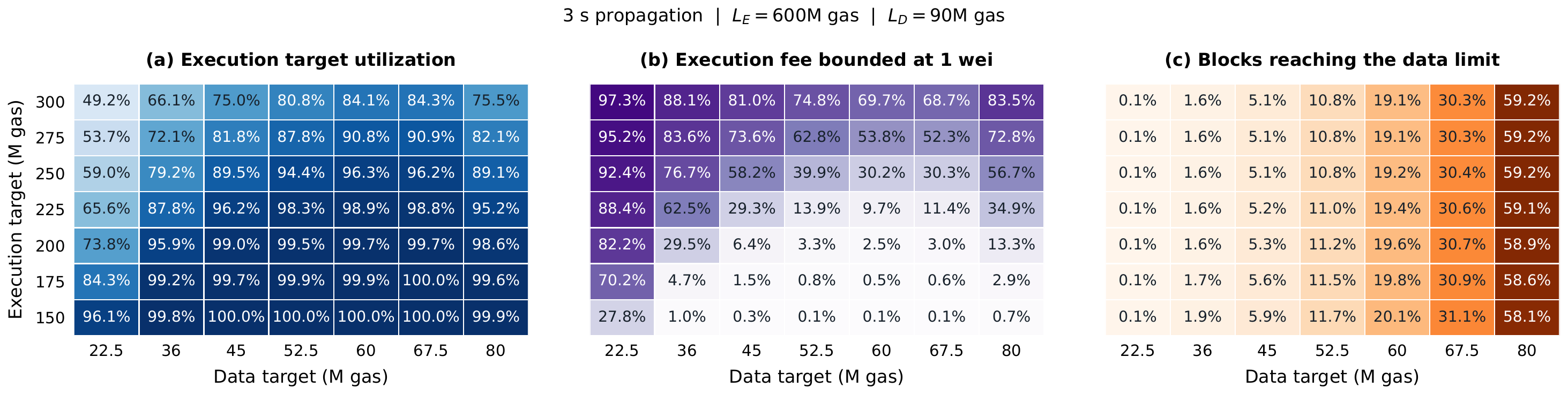}
\caption{(a) Execution target utilization, execution gas usage divided by the target. (b) Fraction of blocks where the execution fee is bounded at 1 wei while gas usage is below target. (c) Fraction of blocks whose data gas usage reaches the data limit.}     
\label{fig:7999-heatmap}
\end{figure*}

The frontier above is bounded. With the data fee at the 1-wei minimum, the lowest effective execution price is $(m_E+w_E)b_{\min}$, so no execution target above
\begin{equation}
    T_E^{\max}=m_E q_E^0\left(\frac{p^0}{(m_E+w_E)\,b_{\min}}\right)^{\epsilon_E}
    \label{eq:execution-ceiling}
\end{equation}
can clear at any data target or propagation allocation. Under the central calibration, $T_E^{\max}\approx 309$M. The required data target rises steeply toward this ceiling, from 57.4M at a 250M execution target to about 144M at the ceiling, where the data fee reaches 1~wei. The frontier involves only targets, so it is the same across propagation allocations. Limits restrict which target pairs can be configured. 

\subsection{Utilization under fluctuating demand}
\label{subsec:7999-fixed-capacity}

\parhead{Fixed capacity with 3-second propagation}
We first fix the propagation allocation at 3 seconds, giving $L_E=600$M and $L_D=90$M, and vary the execution and data targets. We denote by "E250/D60" a configuration with a 250M execution target and a 60M data target. We test seven execution targets from 150M to 300M, 1/4 to 1/2 of the 600M limit, along with seven data targets from 22.5M to 80M, 1/4 to 8/9 of the 90M limit, and evaluate all configurations under the same empirical demand paths as the one-dimensional benchmarks. \Cref{fig:7999-heatmap} reports the simulation outcomes.
 
At each execution target, execution target utilization in \Cref{fig:7999-heatmap}(a) first rises with the data target and then falls as the data target approaches the 90M limit. In \Cref{fig:7999-heatmap}(b), the fraction of blocks in which the execution fee is bounded at 1 wei mirrors this pattern: where utilization is low, the fee is more often bounded. Both effects strengthen with the execution target: at E275, utilization rises from $53.7\%$ at D22.5 to $90.9\%$ at D67.5 before falling to $82.1\%$ at D80.

The two sides of this pattern have different causes, which \Cref{fig:7999-heatmap} panels (b) and (c) separate. At low data targets, execution is constrained by the \emph{data price}. A low data target requires a high data fee to contract static content, and the same fee raises the \gls{bal}-inclusive execution price. Consequently, the execution fee falls to compensate. Once it reaches 1 wei, it can fall no further and execution underfills its target. At E275/D22.5 the execution fee is bounded at 1 wei in 95.2\% of blocks while the data limit binds in only 0.1\%, so the constraint is price rather than capacity.

At high data targets, execution is constrained by the \emph{data limit}. When a positive demand shock in a block pushes demanded data gas above the limit, the block cannot include more \gls{bal}-generating execution because of bundle inclusion. For illustration, E275/D80 clears all targets in equilibrium, with an execution fee of 1.85~wei before integer rounding. However, the data limit binds in 59.2\% of blocks, bundle inclusion excludes the execution that generates the excess \glspl{bal}, and execution averages 225.7M, or 82.1\% of its target, although the execution limit binds in only 0.08\% of blocks. The decline from D67.5 reflects the data limit rather than the data price, since raising the data target lowers the data fee. Therefore, feasibility in equilibrium does not guarantee enough data capacity to support execution when demand varies block by block.

These constraints narrow the feasible data targets as the execution target rises. At a 275M execution target, no data target brings the share of blocks with the execution fee at its 1-wei minimum below $52.3\%$. With a 300M execution target, the frontier data target of 97M exceeds the limit, and the execution fee is bounded at the minimum in at least $68.7\%$ of blocks across the tested grid, limiting execution utilization to at most $84.3\%$. Data-limit pressure itself depends almost entirely on the data target: each column of \Cref{fig:7999-heatmap}(c) varies by about one percentage point across execution targets, rising to roughly $59\%$ at D80.

\begin{figure}[t]
    \centering
    \includegraphics[width=1\linewidth]{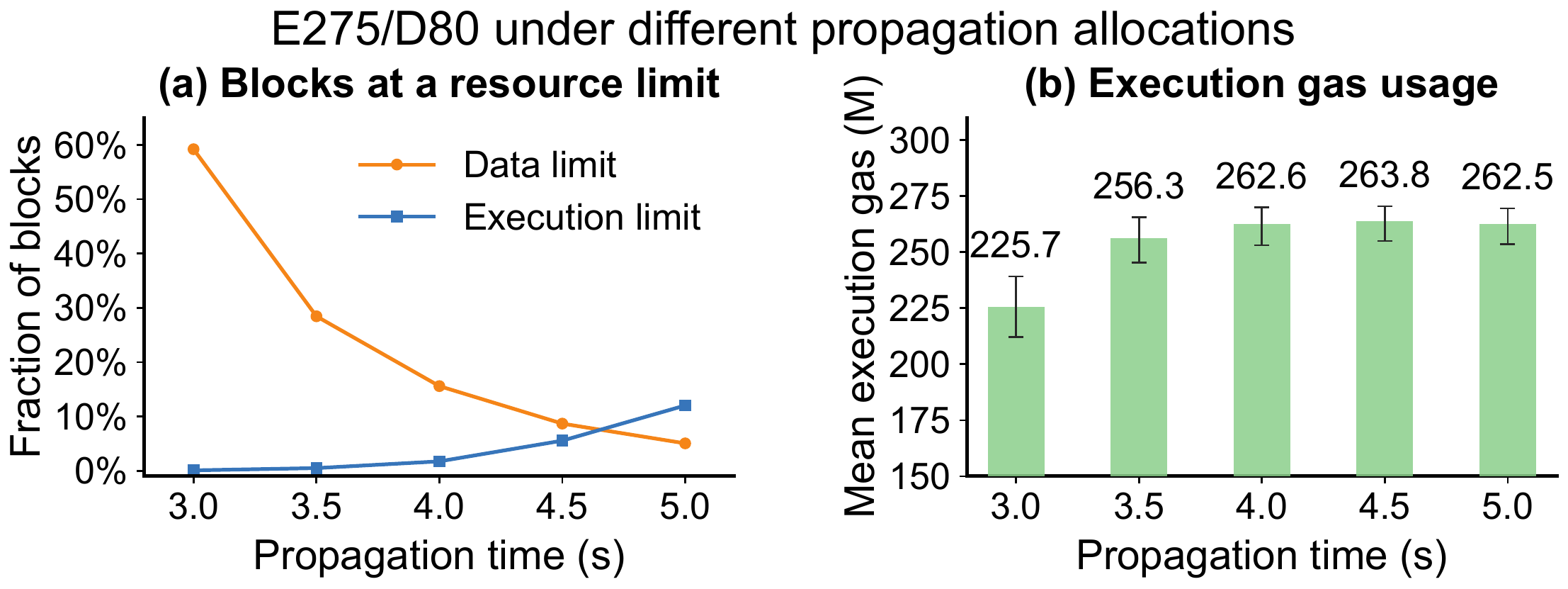}
    \caption{E275/D80 metrics: (a) blocks at each limit,
    (b) mean execution gas, with 5th--95th percentiles across paths.}
    \label{fig:7999-propagation}
\end{figure}

\parhead{Varying propagation time}
At a 3-second allocation, configurations with high execution and data targets lose execution to the 90M data limit. Lengthening propagation raises the data limit at the cost of execution time (cf. \Cref{tab:shared-capacity}), so we ask if additional data capacity recovers execution usage. We simulate the target grid at each allocation, with both limits following it and data targets of 90M and 100M added where the limit allows (cf. \Cref{appendix:comparison}).


For clarity, \Cref{fig:7999-propagation} traces E275/D80, the configuration that illustrated bundle exclusion above. Other configurations with high execution and data targets follow the same pattern. Longer propagation makes the data limit bind less often, and execution rises as less \gls{bal}-generating execution is excluded, even though the extra capacity also raises the data fee. At 3 seconds, the limit sits only 10M above the target, so the fee averages 5~wei, whereas at 4 seconds it averages 43.6~wei, lowering demanded execution from 307.7M to 280.7M through the \gls{bal} charge. Less exclusion also lets the execution fee track demand: at 3 seconds, exclusion keeps usage below target despite excess demand, pushing the fee to 1~wei in $72.8\%$ of blocks, against $36.1\%$ at 4 seconds. 


The remaining shortfall reflects how little room the equilibrium fee leaves. At continuous equilibrium fees of 1.85~wei for execution and 19.0~wei for data, lowering the execution fee to the minimum cuts execution's effective price by only about $31\%$, since the \gls{bal} charge does not fall, raising inelastic execution demand by only about $4.5\%$. Larger negative execution demand shocks and data-fee spikes can leave the execution fee at 1 wei, and execution usage reaches $95.5\%$ of its target at 4 seconds.

Beyond 4 seconds, the execution time also caps further gains. The execution limit falls to 450M and 400M, and its binding frequency rises to $5.5\%$ and $12.0\%$, overtaking the data limit after 4.5 seconds. Execution usage plateaus around 263M, as the remaining data-limit relief and growing execution-limit pressure roughly offset. An equilibrium execution fee above the minimum is thus necessary but not sufficient: a target pair can clear in equilibrium while sitting one modest shock from the minimum, which motivates the room requirement in the following configuration selection.


\section{Comparison}
\label{sec:comparison}

We compare the multi-dimensional design with one-dimensional benchmarks on the same demand paths, reporting mean execution, annualized state growth, and congestion, measured as the share of \emph{near-limit} blocks, whose gas usage reaches at least $98\%$ of any limit. Since the execution and data targets of \acrshort{eip}-7999 are design parameters, we select one configuration per propagation allocation in two ways. \emph{Maximum throughput} takes the highest mean execution, regardless of how often the execution fee sits at its minimum or how congested blocks are. \emph{Conservative} takes the highest-execution configuration that lies inside the execution--data frontier, so that its execution fee has room to fall, and whose near-limit share does not exceed the historical $4.73\%$ over February--May 2026. Conservative configurations select execution targets of 175--225M, and maximum-throughput configurations all select a 300M execution target. We compare them with the normalized one-dimensional benchmark at target ratios of 1/2 and 4/5. \Cref{appendix:comparison} lists all selected configurations, metrics, and path-by-path results.

\begin{figure}[t]
    \centering
    \includegraphics[width=1\linewidth]{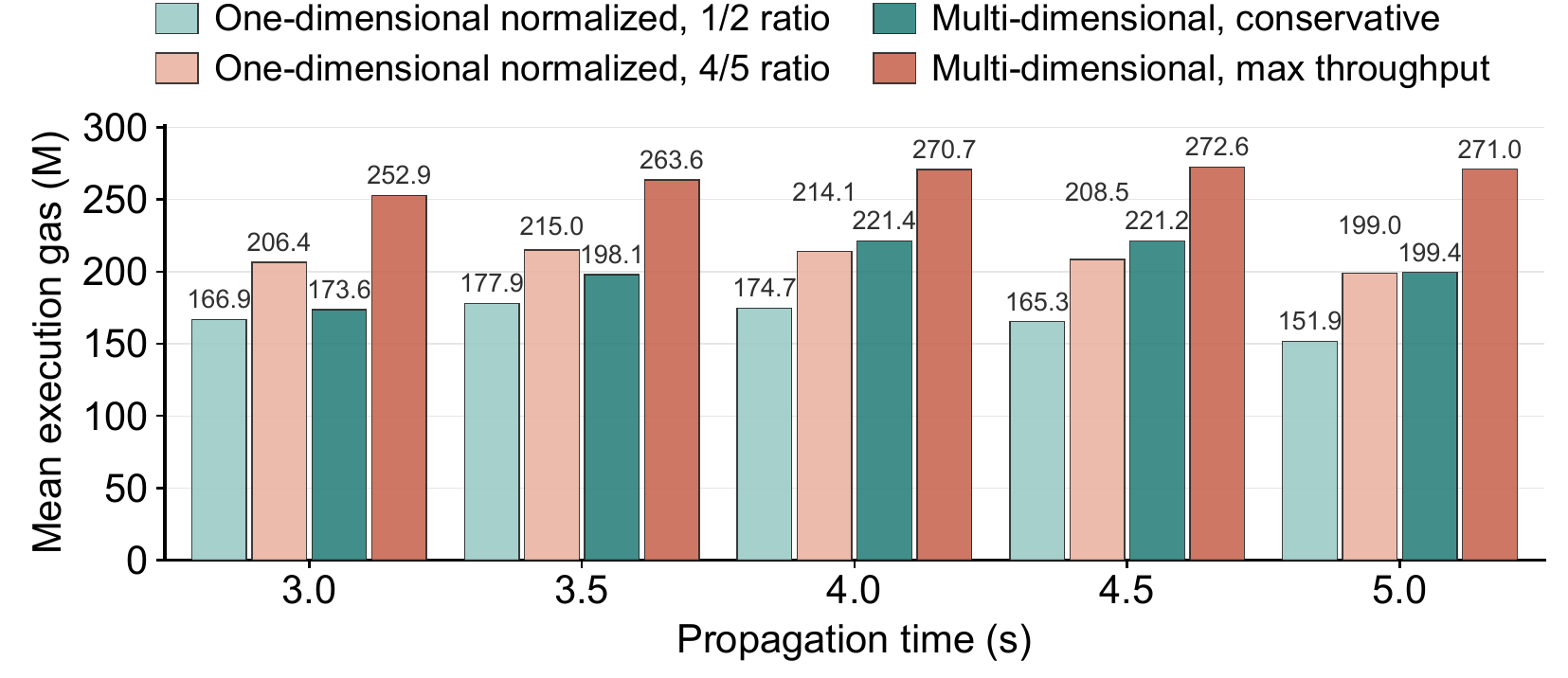}
    \caption{Mean execution by propagation allocation. Colors distinguish congestion levels. Lighter and darker shades denote one- and multi-dimensional configurations.}
\label{fig:comparison}
\vspace{-0.2cm}
\end{figure}

\Cref{fig:comparison} compares the two mechanisms at two levels of congestion. At low congestion, conservative multi-dimensional configurations, with $1.9\%$--$4.4\%$ near-limit blocks, deliver 6.8--56.0M more execution on average than the normalized benchmark at a 1/2 target ratio, with $4.0\%$--$4.5\%$, at every allocation. The variation in gains reflects changes in both the selected multi-dimensional configurations and the one-dimensional benchmark's outcomes. 
We note that allowing $25\%$ more near-limit blocks raises this gain further to 31.2--57.7M. At high congestion, maximum-throughput configurations exceed the normalized benchmark at a 4/5 target ratio by 46.5--72.0M at every allocation, though with more near-limit blocks, $16.6\%$--$32.1\%$ against $13.2\%$--$20.6\%$, and with the execution fee bounded by the minimum in $54.3\%$--$68.7\%$ of blocks. The normalized benchmark at a 4/5 target ratio exceeds the conservative configurations at the 3- and 3.5-second allocations, but only with far more congestion.

Multi-dimensional configurations also hold state growth near its 120~GiB/year target. Among the one-dimensional benchmarks, only the growth-matched one comes close, at 121.4--121.6~GiB/year, by raising the base fee until execution falls to 67.5--70.2M (cf. \Cref{tab:one-dim-outcomes}). Replacing the \gls{eip}-1559 update rule with an \gls{eip}-4844-style fake-exponential update raises the normalized benchmark at a 4/5 target ratio to 244.2M execution and 94.0--96.0~GiB/year of state growth, but both branches still underfill their targets. Execution stays below maximum throughput at every allocation, and about $40\%$ of blocks are near a limit. None of the one-dimensional benchmarks we test keeps state growth near its target while supporting high execution. Thus, in the two main comparisons, the selected multi-dimensional configurations achieve higher target utilization in both dimensions: execution versus regular gas, and state versus normalized state gas.
\section{Discussion and Conclusion}
\label{sec:conclusion}

\parhead{Limitations}
We model demand as independent across resources because Ethereum's resources have always shared one fee, and historical capacity increases identify each resource's response to a common price change. The central calibration uses five-week windows, and our results hold
qualitatively across alternative calibrations, although execution throughput levels depend on the calibration (cf. \Cref{appendix:elasticity_sensitivity}). The main uncertainty is extrapolation. The tested execution targets are 7--9 times the metered anchor, and whether demand of that size materializes, with its composition evolving as the calibrated elasticities imply, cannot be established from historical data. Thus, our results are conditional on the calibrated demand.

Since \glspl{bal} do not exist historically, their demand response cannot be estimated. We fold their floor uplift into static-content demand under the one-dimensional benchmarks, an approximation weaker at high floor rates (cf. \Cref{appendix:data_metering}), and apply the parent activity's elasticity to its \gls{bal}-inclusive price under the multi-dimensional mechanism.

We do not model the blob fee market. Both mechanisms imply per-byte data charges far below historical blob prices~\cite{hildobby_blob_dash}, and we omit the blob-linked reserve price of EIP-7999~\cite{eip7999pull}. How data and blob prices should interact, and whether protocol-generated \glspl{bal} should face a reserve, warrant further study. The costs of transitioning to a multi-dimensional mechanism, including changes to transaction-level gas limits, \gls{evm} gas accounting, and user fee estimation, also remain unquantified.

\parhead{Conclusion}
Under calibrated demand, state, the most price-responsive and heavily metered resource, sets the fee under the one-dimensional mechanism, suppressing execution. State normalization aligns the regular and state branches under fixed demand, but a single fee cannot track both as demand varies. The tested fake-exponential update recovers execution, but with much higher congestion. At comparable congestion, selected multi-dimensional configurations deliver 4--34\% more execution across propagation allocations, with higher target utilization and state growth near its target. These results support separate pricing as Ethereum scales, with data targets and limits accounting for execution-generated data and execution targets within the execution--data frontier.

More broadly, other blockchains face the same choice between adding meters and adding prices: Aptos charges for persistent storage separately and has explored multi-dimensional gas markets~\cite{goren2024multidimensional}, and Solana has proposed separately metering account-state loading~\cite{solana_chili_peppers}. Our results suggest two lessons. First, separate meters under a single fee still allow the bottleneck resource to determine the price paid by other activities, and recalibrating the meter aligns the resources only under a particular demand mix. Second, separate prices may not remove resource dependencies: an activity that generates demand for another resource can remain constrained by that resource's price and capacity.

\section*{Use of Generative AI} 
The manuscript was primarily drafted by the authors. We used ChatGPT and Claude to assist with discussing and refining the model, checking mathematical derivations, developing simulation code, and shortening selected passages. The authors made the final methodological and interpretive decisions and take full responsibility for the content of this paper.

\bibliographystyle{ACM-Reference-Format}
\bibliography{sample-base}

\appendix
\crefalias{section}{appendix}
\crefalias{subsection}{appendix}
\crefalias{subsubsection}{appendix} 
\section{Metering and calibration}
\label{appendix:metering_calibration}

\begin{figure*}[t]
\centering
\includegraphics[width=\textwidth]{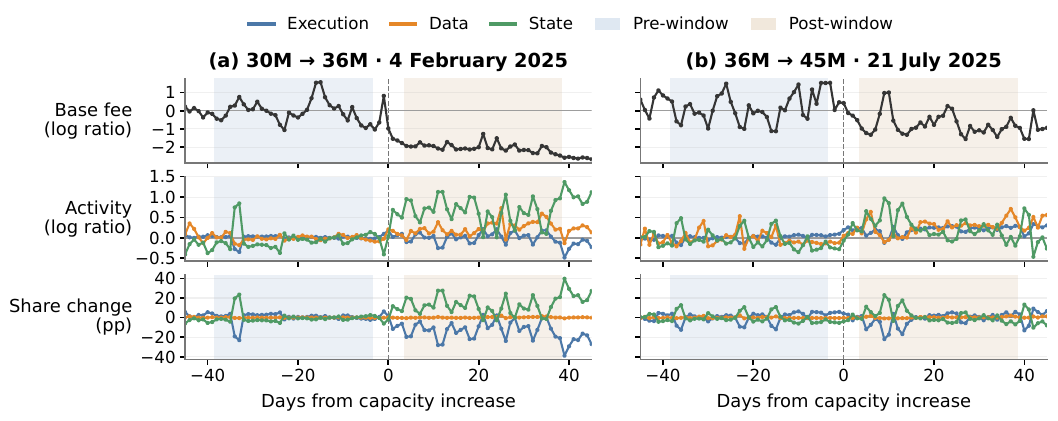}
\caption{Observed daily fees, resource activity, and shares around the two calibration events. Shading marks the pre/post windows. Fees and activity are log ratios to pre-period references, and gas shares are percentage-point differences.}
\label{fig:calibration-event-time}
\end{figure*}

\parhead{Elasticity calibration}
\label{appendix:elasticity_calibration}
For each retained event $e\in\{1,2\}$, let superscripts $-$ and $+$ denote the pre- and post-event windows. The responses of resource activity and total gas are
\begin{equation}
    r_{i,e}=-\frac{\ln(q_{i,e}^{+}/q_{i,e}^{-})}{\ln(p_e^{+}/p_e^{-})},
    \qquad
    a_e=-\frac{\ln(Q_e^{+}/Q_e^{-})}{\ln(p_e^{+}/p_e^{-})},
    \label{eq:event-responses}
\end{equation}
where $Q_e^{\pm}=\sum_i q_{i,e}^{\pm}$. We calibrate
\begin{equation}
    \epsilon_i=\frac{1}{2}\sum_{e=1}^{2}
    \Bigl(r_{i,e}+a_e-\sum_{j\in\Rset}s_j^\star r_{j,e}\Bigr),
    \label{eq:elasticity-estimator}
\end{equation}
where $s_j^\star$ is resource $j$'s mean daily share of gas during the 45M-gas regime (21 July to 24 November 2025): $s_\rE^\star=0.757$, $s_\rD^\star=0.036$, and $s_\rS^\star=0.207$. Because the shares sum to one, $\sum_j s_j^\star\epsilon_j$, the model's elasticity of total gas at these shares, equals the average observed total-gas response, while the differences between resource elasticities equal the average differences between event responses. The resulting shift is common to all resources: 0.004 under the central calibration. Matching a finite-change aggregate response to the local elasticity at fixed shares is an approximation, and the share anchor is a calibration choice. Using the 2026 activity shares of \Cref{sec:shared-metering} instead would raise each elasticity by about 0.007, well within the range spanned by the window sensitivity in \Cref{appendix:elasticity_sensitivity}.

\Cref{fig:calibration-event-time} shows the daily fees and resource activity behind these responses. The two events respond differently. In February, the fee begins falling before the increase, and post-event state activity rises by $97.6\%$, while execution rises by only $2.3\%$, consistent with its low event response. In July, execution rises more steadily ($21.8\%$), and state growth is concentrated early in the post-period. These patterns describe the event responses; without an untreated control, they cannot separate capacity effects from contemporaneous changes in demand.

\parhead{Execution metering}
We reconstruct execution gas under \gls{eip}-8038~\cite{eip8038} and
\gls{eip}-2780~\cite{eip2780} using Xatu block, transaction, and
execution-trace data~\cite{xatu}. The panel covers 860,505 mainnet blocks from 1 February to 31 May 2026 (blocks 24,358,293--25,218,797), supplemented with access-list and authorization data from an Erigon archive node. Starting from the historical anchor of 23.942M gas per block, the \gls{eip}-8038 SSTORE increase adds 11.618M, other repricing changes add 0.515M, \gls{eip}-2780's intrinsic-gas changes subtract 1.033M, and refunds under the updated schedule subtract 0.375M, giving 34.667M and $m_\rE=34.667/23.942\approx1.448$.


\parhead{Data metering in one-dimensional benchmarks}
\label{appendix:data_metering}
We uniformly sample 50 blocks per day from the same panel with a fixed random seed, yielding 6,000 blocks containing 1,899,748 transactions. We retrieve complete transaction bodies from an Erigon archive node to measure calldata, access lists, authorization tuples, and blob versioned
hashes~\cite{eip8131}. Combining these contents with Xatu gas usage and
execution records, we reconstruct the transaction floor, including runtime \gls{bal} charges~\cite{eip8279}, at each specified floor rate.
\Cref{tab:data-multiplier} reports the resulting data multipliers.

\begin{table}[h]
    \centering
    \small
    \setlength{\tabcolsep}{2pt}
    \begin{tabular}{@{}>{\raggedright\arraybackslash}p{.45\columnwidth}rrrrr@{}}
    \toprule
    Accounting step & $F=40$ & $F=50$ & $F=64$ & $F=82$ & $F=96$\\
    \midrule
    \multicolumn{6}{@{}l}{\emph{Data-attributed gas per block} (M gas)}\\
    \quad Current anchor
        & 1.181 & 1.181 & 1.181 & 1.181 & 1.181\\
    \quad $+$ static-content floor~\cite{eip8131}
        & 1.830 & 2.000 & 2.335 & 2.988 & 3.654\\
    \quad $+$ runtime \gls{bal} floor~\cite{eip8279}
        & 1.876 & 2.082 & 2.552 & 3.645 & 4.761\\
    \midrule
    $m_\rD^{\mathrm{floor}}(F)$
        & 1.589 & 1.763 & 2.162 & 3.087 & 4.033\\
    \addlinespace
    Transactions \gls{bal}-repriced
        & 0.5\% & 0.8\% & 2.0\% & 4.5\% & 6.8\%\\
    \bottomrule
    \end{tabular}
    \caption{Mean data-attributed regular gas after each accounting step. The last row counts transactions whose regular gas changes when \gls{bal} enters the floor.}
    \label{tab:data-multiplier}
\end{table}

\parhead{\gls{bal}-related floor exposure}
A cold storage read adds $32F$ floor gas through its \gls{bal}, exceeding its 2,100-gas execution charge when $F>65.625$~\cite{eip8038}. At $F=82$ and $F=96$, access-intensive transactions with little execution headroom can become floor-bound through \glspl{bal} even when their static-content floor does not bind. In the sample, \glspl{bal} change regular gas for $4.46\%$ and $6.79\%$ of transactions at these two rates, accounting for $1.68\%$ and $2.75\%$ of \gls{bal}-inclusive regular gas. Most of this uplift, $71.0\%$ and $62.5\%$, comes from transactions already bound by the static floor or with at most $10\%$ execution headroom, while newly floor-bound transactions contribute $21.3\%$ and $31.7\%$. Thus, we fold the uplift into $m_\rD^{\mathrm{floor}}(F)$ as an aggregate approximation.
Assigning its demand response to static-data activity remains an assumption, since historical data do not identify responses to these charges.

\parhead{\gls{bal} attribution}
We reconstruct runtime \gls{bal} bytes under \gls{eip}-8279~\cite{eip8279} for the same 6,000 blocks using Xatu opcode aggregates, execution traces, storage diffs, and contract and account records~\cite{xatu}. New storage values and deployed code are attributed directly to state creation. For keys, addresses, and balances, we estimate creation-related bytes from each transaction's new-state counts, capped by its corresponding runtime events, and attribute the remaining bytes to execution and state access, including access performed by state-creating transactions. \Cref{tab:bal-decomposition} reports the result. At 16 gas per byte, mean \gls{bal} gas per block is 1.919M, giving $w_S=0.0417$ and $w_E=0.0710$.

\begin{table}[h]
\centering
\small
\setlength{\tabcolsep}{5pt}
\begin{tabular}{@{}lrr@{}}
\toprule
Runtime component & State creation & Execution\\
\midrule
Storage key & 5,523 & 71,235\\
Storage value & 5,591 & 22,538\\
Account access & 162 & 8,360\\
Value transfer & 633 & 4,135\\
Contract creation & 200 & 9\\
Deployed code & 1,555 & 0\\
\midrule
Total & 13,666 & 106,278\\
Share & 11.4\% & 88.6\%\\
\bottomrule
\end{tabular}
\caption{Mean runtime \gls{bal} bytes per block attributed to state creation
and execution in the 6,000-block sample.}
\label{tab:bal-decomposition}
\end{table}

\parhead{One-dimensional benchmark parameters}
\label{appendix:one_dimensional}
Combining these coefficients with the capacities in \Cref{tab:shared-capacity}
gives the calibrated parameters of the growth-matched and normalized one-dimensional benchmarks in
\Cref{tab:one-dimensional-calibration}. The baseline keeps $\cpsb=1530$ and $m_\rS=5.656$. The large \cpsb~and state metering coefficient in the normalized benchmark compensate for lower fees, so they do not imply a comparable increase in the ETH price of a state byte in equilibrium.

\begin{table}[h]
    \centering
    \footnotesize
    \setlength{\tabcolsep}{3pt}
    \renewcommand{\arraystretch}{0.95}
    \begin{tabular}{@{}lrrrccrr@{}}
    \toprule
    Benchmark & $\tau$ (s) & $L_G$ (gas) & $F$
        & $m_\rD^{\mathrm{floor}}(F)$ & CPSB & $m_\rS$ & $k$\\
    \midrule
    \multirow{5}{*}{\shortstack[l]{Growth-\\matched}} & 3.0 & 534.0M & 96 & 4.033 & 5,446 & 20.135 & \textemdash{}\\
     & 3.5 & 550.0M & 82 & 3.087 & 5,610 & 20.740 & \textemdash{}\\
     & 4.0 & 500.0M & 64 & 2.162 & 5,100 & 18.854 & \textemdash{}\\
     & 4.5 & 450.0M & 50 & 1.763 & 4,590 & 16.969 & \textemdash{}\\
     & 5.0 & 400.0M & 40 & 1.589 & 4,080 & 15.084 & \textemdash{}\\
    \midrule
    \multirow{5}{*}{Normalized} & 3.0 & 534.0M & 96 & 4.033 & 20,538,083 & 75,928.021 & 3,771.22\\
     & 3.5 & 550.0M & 82 & 3.087 & 37,466,671 & 138,511.963 & 6,678.55\\
     & 4.0 & 500.0M & 64 & 2.162 & 32,147,026 & 118,845.565 & 6,303.33\\
     & 4.5 & 450.0M & 50 & 1.763 & 19,653,155 & 72,656.497 & 4,281.73\\
     & 5.0 & 400.0M & 40 & 1.589 & 9,512,074 & 35,165.549 & 2,331.39\\
    \bottomrule
    \end{tabular}
    \caption{Calibration of the one-dimensional benchmarks under each
    propagation allocation.}
    \label{tab:one-dimensional-calibration}
\end{table}

\section{Demand path construction}
\label{appendix:demand_paths}


We construct demand paths from Ethereum mainnet activity collected through Xatu~\cite{xatu}, using 430,605 consecutive blocks from April--May 2026 (blocks 24,788,193--25,218,797), together with reconstructed runtime \gls{bal}. Observed activity reflects both demand variation and responses to historical fees, and the same execution and state activity can generate different amounts of \gls{bal} data. For each resource $i\in\{E,D,S\}$, we construct a fee-adjusted demand factor and a \gls{bal} composition factor,
\begin{equation}
    \widetilde{s}_{i,t}
    =\frac{q_{i,t}^{\mathrm{obs}}}{q_i^0}
     \Bigl(\frac{p_t}{p^0}\Bigr)^{\epsilon_i},
    \qquad
    \widetilde{a}_t
    =\frac{g_{\mathrm{BAL},t}^{\mathrm{obs}}}
          {w_Eq_{E,t}^{\mathrm{obs}}+w_Sq_{S,t}^{\mathrm{obs}}},
    \label{eq:historical-adjustment}
\end{equation}
where $q_{i,t}^{\mathrm{obs}}$ is historical activity, $p_t$ the historical shared base fee, and $g_{\mathrm{BAL},t}^{\mathrm{obs}}$ observed runtime \gls{bal} gas. The first removes the estimated fee response, so that the simulations do not apply a counterfactual fee response on top of the historical one. The second measures \gls{bal} relative to what its parent activities predict, normalized to preserve the \gls{bal} anchor.

We jointly resample block-level variation in execution, static data, state, and \gls{bal} composition in contiguous 3,200-block segments, with the segment length selected using autocorrelation and burst diagnostics. We combine these draws with jointly sampled eight-day sequences of daily conditions and the historical hourly profiles. This construction retains persistence and co-movement across resources at different timescales. Source components are normalized before sampling, but individual paths are not renormalized, so sampled busy or quiet weeks retain their demand levels. Each simulated block
then receives demand $q_{i,t}=q_i^0s_{i,t}(p_{i,t}/p^0)^{-\epsilon_i}$ at its simulated effective price, with parent prices using average \gls{bal} intensities. We generate 32 paths, each with a one-day burn-in and seven measured days, shared by all mechanisms, configurations, and elasticity calibrations.

\parhead{Block inclusion}
In the multi-dimensional simulations, let $g_{E,t}$ denote block $t$'s demanded execution gas, and $L_E$ and $L_D$ the execution and data limits (cf. \Cref{tab:shared-capacity}). We first cap execution at its limit by the factor $\eta_t=\min\{1,L_E/g_{E,t}\}$, removing its \gls{bal} with it, so that \gls{bal} from excluded execution does not count against the data limit. Let $\widetilde g_{D,t}$ denote the remaining data demand: static content plus the \gls{bal} of included execution and of state. If it exceeds $L_D$, we scale all remaining activity by $\theta_t=\min\{1,L_D/\widetilde g_{D,t}\}$, without modeling transaction-level priorities. Included execution, static data, and state are therefore $\eta_t\theta_tq_{E,t}$, $\theta_tq_{D,t}$, and $\theta_tq_{S,t}$, so a binding data limit reduces execution even when the execution limit is slack. In the one-dimensional simulations, a common factor scales all activity when either branch exceeds its limit.

\section{Comparison details}
\label{appendix:comparison}

\Cref{tab:comparison} reports outcomes for each design at each allocation.

\begin{table*}[!t]
\centering
\small
\setlength{\tabcolsep}{3pt}
\begin{tabular}{@{}lrrrccrccc@{}}
\toprule
& \multicolumn{3}{c}{Elasticity}
& \multicolumn{2}{c}{Minimum $S(b_R)/T_G$}
& & \multicolumn{3}{c}{Multi-dimensional execution gain (M)}\\
\cmidrule(lr){2-4}\cmidrule(lr){5-6}\cmidrule(l){8-10}
\makecell{Window\\(days)} & $\epsilon_E$ & $\epsilon_D$ & $\epsilon_S$
& Baseline & Growth-matched & \makecell{$T_E^{\max}$\\(M)}
& \makecell{Conservative\\vs.\ normalized 1/2}
& \makecell{Maximum\\vs.\ normalized 4/5}
& \makecell{Maximum vs.\\fake exp.\ 4/5}\\
\midrule
21 & 0.117 & 0.202 & 0.478 & 34.5 & 87.9 & 287.5 & 41.7--79.6 & 58.8--78.0 & 50.2--76.7\\
35 & 0.121 & 0.229 & 0.335 & 5.4 & 13.4 & 309.5 & 6.8--56.0 & 46.5--72.0 & 19.5--56.2\\
60 & 0.082 & 0.205 & 0.280 & 8.7 & 22.5 & 151.6 & 10.2--32.1 & 19.5--30.4 & 8.2--23.6\\
75 & 0.079 & 0.201 & 0.254 & 6.7 & 17.0 & 143.2 & 12.7--31.5 & 18.0--27.4 & 6.1--18.9\\
\bottomrule
\end{tabular}
\caption{Robustness to the elasticity calibration. Minimum $S(b_R)/T_G$ is taken across propagation allocations, and the execution ceiling uses $b_E=b_D=1$~wei. Execution gains are ranges across the five allocations, with \gls{eip}-7999 configurations reselected under each calibration and the one-dimensional benchmarks held at their central calibration.}
\label{tab:elasticity-sensitivity}
\end{table*}

\begin{table}[H]
\centering\footnotesize
\setlength{\tabcolsep}{3.5pt}
\begin{tabular}{@{}lrrrrr@{}}
\toprule
& \multicolumn{5}{c}{Propagation allocation $\tau$ (s)}\\
\cmidrule(l){2-6}
& 3.0 & 3.5 & 4.0 & 4.5 & 5.0\\
\midrule
\multicolumn{6}{@{}l}{\emph{Baseline}}\\
Execution (M)   & 82.0  & 87.9  & 92.6  & 89.1  & 85.3\\
State (GiB/yr)  & 288.1 & 349.0 & 403.4 & 362.5 & 321.4\\
Near-limit (\%) & 8.50  & 8.69  & 8.79  & 8.72  & 8.62\\
\addlinespace
\multicolumn{6}{@{}l}{\emph{Growth-matched}}\\
Execution (M)   & 70.0  & 70.2  & 69.4  & 68.5  & 67.5\\
State (GiB/yr)  & 121.6 & 121.6 & 121.6 & 121.5 & 121.4\\
Near-limit (\%) & 9.01  & 9.02  & 9.01  & 8.97  & 8.92\\
\addlinespace
\multicolumn{6}{@{}l}{\emph{Normalized, ratio 1/2}}\\
Execution (M)   & 166.9 & 177.9 & 174.7 & 165.3 & 151.9\\
State (GiB/yr)  & 88.1  & 86.0  & 86.0  & 87.0  & 87.8\\
Near-limit (\%) & 4.48  & 4.05  & 3.98  & 4.14  & 4.30\\
\addlinespace
\multicolumn{6}{@{}l}{\emph{Normalized, ratio 4/5}}\\
Execution (M)   & 206.4 & 215.0 & 214.1 & 208.5 & 199.0\\
State (GiB/yr)  & 64.6  & 56.8  & 56.7  & 61.1  & 68.7\\
Near-limit (\%) & 17.73 & 13.16 & 13.37 & 16.02 & 20.62\\
\addlinespace
\multicolumn{6}{@{}l}{\emph{Normalized, ratio 4/5, fake-exponential update}}\\
Execution (M)   & 225.9 & 244.2 & 243.6 & 232.5 & 214.8\\
State (GiB/yr)  & 96.0  & 94.0  & 94.0  & 95.2  & 95.6\\
Near-limit (\%) & 40.76 & 39.07 & 38.84 & 39.64 & 39.85\\
\midrule
\multicolumn{6}{@{}l}{\emph{\gls{eip}-7999, conservative}}\\
Targets E/D (M) & 175/36 & 200/45 & 225/52.5 & 225/52.5 & 200/60\\
Execution (M)   & 173.6  & 198.1  & 221.4    & 221.2    & 199.4\\
Near-limit (\%) & 1.92   & 2.42   & 3.00     & 3.99     & 4.39\\
\addlinespace
\multicolumn{6}{@{}l}{\emph{\gls{eip}-7999, maximum throughput}}\\
Targets E/D (M) & 300/67.5 & 300/67.5 & 300/80 & 300/90 & 300/90\\
Execution (M)   & 252.9 & 263.6 & 270.7 & 272.6 & 271.0\\
Near-limit (\%) & 32.14 & 16.56 & 18.82 & 20.65 & 23.02\\
\bottomrule
\end{tabular}
\caption{Outcomes of the compared designs under the same 32 demand paths and
central elasticities. \gls{eip}-7999 holds state growth at 120.0~GiB/year in every configuration.}
\label{tab:comparison}
\end{table}

\parhead{Fake-exponential update}
\label{appendix:update-rule}
To test whether the one-dimensional shortfall stems from fee rounding, we replace the \gls{eip}-1559 update of the normalized benchmark at a target ratio of 4/5 with the persistent excess-gas counter and fake
exponential of \gls{eip}-4844~\cite{eip4844}, keeping the configurations and 32 demand paths fixed. The counter accumulates small shortfalls instead of discarding them when the fee change rounds to zero. The denominator $8T_G$ matches the response of \gls{eip}-1559. Using $T_G/\ln(1.125)$ instead changes execution by less than 0.014M at every allocation.

In \Cref{tab:comparison}, the fake-exponential update increases execution at every allocation. At 3.5 seconds, execution rises from 215.0M to 244.2M as the mean fee falls from 18.26 to 3.56~wei, and state growth rises from 56.8 to 94.0~GiB/year. Near-limit blocks rise from $13.16\%$ to $39.07\%$, because the 4/5 ratio leaves little headroom for bursts and the lower fee admits more activity. Regular and normalized state utilization remain at $85.8\%$ and $78.4\%$. Fee rounding thus explains part of the shortfall, but the shared-price coupling remains.

\parhead{Paired-path comparison} 
Under the central calibration, we fix each allocation's selected configurations and compare mean included execution gas per block path by path across the same 32 demand paths. Against the normalized benchmark at a ratio of 1/2, the conservative selection gains 6.8--56.0M on average across allocations. It delivers more execution on 30 of 32 paths at 3 seconds and on every path at the other allocations, and the minimum path-level difference is -3.6M. Against the normalized benchmark at 4/5, the maximum-throughput configurations gain 46.5--72.0M on average and at least 35.6M on any path. Against its fake-exponential variant, the corresponding gains are 19.5--56.2M and at least 4.8M. Both maximum-throughput comparisons are positive on every path at every
allocation. Because the configurations were selected on these same paths, this
is an in-sample comparison.

\section{Robustness to elasticity calibration}
\label{appendix:elasticity_sensitivity}

We repeat the analyses under the elasticities from the 21-, 60-, and 75-day calibration windows, holding the activity anchors, metering, physical limits under each propagation allocation, \gls{bal} attribution, and the demand paths fixed.
\Cref{tab:elasticity-sensitivity} summarizes the results.
 
The one-dimensional benchmarks keep their central calibration of \cpsb\ and $k$. These are constants set by the protocol from the gas target and demand estimates, so holding them fixed asks how each benchmark performs when true demand differs from the calibration it was built on. For the multi-dimensional mechanism, we reselect execution and data targets from the tested grid with the same criteria, extending the grid where the targets cannot clear.

\parhead{One-dimensional mechanisms}
Under every calibration, state is the equilibrium bottleneck in the baseline and growth-matched benchmarks: at the regular-clearing fee, state demand exceeds the target at least 5.4-fold and 13.4-fold, respectively, and in simulation state sets the shared fee in at least $92.5\%$ and $97.7\%$ of blocks. Under normalization, both branches remain below target under every calibration, with either the integer or the fake-exponential update. 

A fixed $k$ aligns the branches only under the demand it was calibrated on. With $k$ at its central value, normalized state demand at the regular-clearing fee is 2.1--2.8 times its target under the other
calibrations, so state again sets the fee, and regular gas falls to 64--81\% of its target even in equilibrium; realignment would require scaling $k$ by a factor of 5--44. Therefore, a one-time calibration of $k$, as specified in \cite{eip8372}, may not be sufficient if demand changes.

\parhead{Multi-dimensional mechanism}
The findings of \Cref{sec:7999} hold under every calibration. The execution--data frontier keeps its shape, with the required data target rising steeply toward the execution ceiling, but it shifts with that ceiling. Under the 60- and 75-day calibrations, the lower execution elasticity reduces $T_E^{\max}$ to 151.6M and 143.2M, so nearly all of the tested grid cannot clear in equilibrium. Therefore, we extend the grid to execution targets of 100--140M. Within the extended grid, utilization at a fixed execution target first rises and then falls with the data target, and longer propagation first relieves the data limit, although the allocation at which execution peaks shifts.

\parhead{Comparison} 
After reselection, findings in \Cref{sec:comparison} hold under every calibration. The last three columns of \Cref{tab:elasticity-sensitivity} show that the selected multi-dimensional configurations deliver more execution than the normalized benchmark at every allocation, both at low and at high congestion and against the fake-exponential update, while holding state growth near its target. The gains are smaller under the 60- and 75-day calibrations, where lower execution elasticities limit how much execution any mechanism can attract. 
\end{document}